\documentclass{appolb}
\usepackage{amsmath}
\usepackage{amssymb}
\usepackage{epsfig}

\begin{document}
\title{Lattice searches for tetraquarks and mesonic molecules: light scalar mesons and XYZ states
\thanks{Presented at Excited QCD 2010, 31 January-6 February 2010, Tatra National Park, Slovakia}%
}
\author{Sasa Prelovsek
\address{Department of Physics, University of Ljubljana and Jozef Stefan Institute, Ljubljana, Slovenia}
}
\maketitle
\begin{abstract}
Searches for tetraquarks and mesonic molecules in lattice QCD are briefly reviewed. In the light quark sector the most serious candidates are the lightest scalar resonances $\sigma$, $\kappa$, $a_0$ and $f_0$. In the hidden-charm sector I discuss lattice simulations  of $X(3872)$, $Y(4260)$, $Y(4140)$ and $Z^+(4430)$. The most serious challenge in 
all these lattice studies is the presence of scattering states in addition to possible tetraquark/molecular states. The available  methods for distinguishing both are   reviewed and the main conclusions of the simulations are presented.      
\end{abstract}
  
\section{Introduction}

Some of the observed resonances, i.e. light scalars \cite{tetra_scalar_phenomenology} and some hidden-charm resonances \cite{swanson},  are strong candidates for tetraquarks $[qq][\bar q\bar q]$ or mesonic molecules $(\bar qq)(\bar qq)$. Current lattice methods do not distinguish between both types, so a common name ``tetraquarks'' will be often used to denote both types of $\bar q\bar qqq$ Fock components below.  

In order to extract the information about tetraquark states,  lattice QCD simulations evaluate correlation functions on $L^3\times T$ lattice with tetraquark interpolators ${\cal O}\sim \bar q\bar qqq$ at the  source and the sink
\begin{equation}
\label{cor}
C_{ij}(t)=\langle 0| {\cal O}_i(t){\cal O}^{\dagger}_j(0)|0\rangle_{\vec p =\vec 0}\stackrel{T\to \infty}{\longrightarrow} \sum_n Z_i^{n}Z_j^{n*}~e^{-E_n~t}~\quad n=1,2,\cdots
\end{equation} 
If the  correlation matrix is calculated for a number of interpolators ${\cal O}_{i=1,..,N}$ with given quantum numbers, the energies of the few lowest physical states $E_n$ and the corresponding couplings $Z_i^{n}\equiv\langle 0| {\cal O}_i|n\rangle$
can be extracted from the eigenvalues $\lambda^n(t)=e^{-E_n(t-t_0)}$ and eigenvectors $\vec u^n(t)$ of the generalized eigenvalue problem $C(t)\vec u^{n}(t)=\lambda^{n}(t,t_0) C(t_0)\vec u^{n}(t)$, as discussed in \cite{var_alpha}. 

In addition to possible tetraquarks, also the two-meson scattering states $M_1M_2$ unavoidably contribute to the correlation function and this presents the main obstacle in extracting the information about tetraquarks. The scattering states $M_1(k)M_2(-k)$ at total momentum $\vec p=\vec 0$ have discrete energy levels  
\begin{equation}
\label{enM}
E_{M_1M_2}\simeq E_{M_1}(k)+E_{M_2}(-k)~,\quad E_M(k)=\sqrt{m_M^2+\vec k^2}~,\quad \vec k=\tfrac{2\pi}{L}\vec n
\end{equation}    
in the non-interacting approximation when periodic boundary conditions in space are employed. 

The resonance manifests itself on the lattice as a state in addition to the discrete tower of scattering states (\ref{enM}) \cite{tetra_dyn_proc,tetra_dyn,mathur_scalar} and it is often above the lowest scattering state  (at $E\simeq M_1+M_2$ for S-wave decay). 
 So the extraction of a few  states in addition to the ground state may be crucial. However, many simulations extract only the ground state energy $E_1$ using a conventional exponential fit  $\langle 0| {\cal O}(t){\cal O}^{\dagger}(0)|0\rangle \propto e^{-E_1 t}$ at large $t$. 

Once the physical states are obtained, 
 one needs to determine whether a certain state corresponds to a one-particle (tetraquark)  or a two-particle (scattering) state and the available methods to distinguish both are reviewed in the next Section. 

\section{Methods to distinguish one-particle and scattering states}
I am listing the available methods, which may be complementary:

\begin{itemize}
\item For a one-particle state $n$ the coupling $Z_i^n$ is expected to be almost independent of the lattice size $L$, i.e. $Z_i^n(L_1)/Z_i^n(L_2)\simeq 1$.  For a two-particle state $n=M_1M_2$ one expects  $Z_i^n(L_1)/Z_i^n(L_2)\simeq (L_2/L_1)^{3/2}$ if the range of interaction between $M_1$ and $M_2$ is much smaller than $L$ \cite{tetra_dyn_proc,tetra_dyn,mathur_scalar,hsieh,tetra_sasa}. But this method leads to a reliable distinction only in presence of long stable plateaus, as cautioned in \cite{alexandrou}.    
\item One can distinguish whether the ground state is a one-particle or a two-particle state from the  time-dependence of the $C_{ii}(t)$ near $t\simeq T/2$ at finite temporal extent $T$. In case of  (anti)periodic boundary conditions  $C_{ii}(t)=|Z_i^1|^2~[e^{-E_1t}+\{t\to T-t\}]$ for one-particle ground state and $C_{ii}(t)= |Z_i^{1}|^2~e^{-E_1t}+|\tilde Z_i^{1}|^2~e^{-m_{M_1}t}e^{-m_{M_2}(T-t)}+\{t\to T-t\}$ ($\tilde Z_i^{n} =\langle M_1^\dagger |{\cal O}_i|M_2\rangle$) for two-particle ground state \cite{tetra_sasa}. The criteria for distinguishing the excited states is discussed in \cite{tetra_dyn}.
\item An attractive interaction  between two particles  in a scattering state  manifests itself by a scattering length $a>0$.  
A formation of a  bound state below certain $m_\pi$  can be identified by the change of sign of $a$ from positive to negative as $m_\pi$ is lowered \cite{liuming,Z_chinese}.   The scattering length $a$ for S-wave scattering $M_1M_2$ can be determined from the energy shift $\Delta E=E_1-m_{M_1}-m_{M_2}$ on a finite lattice \cite{liuming,Z_chinese}. 
\item Certain non-conventional spatial boundary conditions have specified effects on the one- and two-particle energies, which allow to distinguish both types \cite{suganuma}. 
\end{itemize}
\section{Light scalar resonances}

It is still not established whether the lightest scalar mesons  $\sigma$,  $\kappa$, $a_0(980)$ and $f_0(980)$  are conventional  $\bar qq$ states or  have an important $\bar q \bar qqq$  Fock component, as strongly supported by some phenomenological studies \cite{tetra_scalar_phenomenology}. The tetraquark interpretation implies that the $I=1$ state 
 ($\bar u\bar s sd$)  is heavier than the $I=1/2$ state ($\bar u\bar d ds$) due to $m_s>m_d$, in agreement with experimental ordering $m_{a_0(980)}>m_{\kappa}$.   One the other hand,  the conventional $\bar ud$ and $\bar us$ states can hardly explain the observed mass ordering. The tetraquark interpretation also naturally explains the large observed coupling of $a_0(980)$ and $f_0(980)$ to $\bar KK$, which is due to the additional valence pair $\bar ss$.

All lattice simulations that look for tetraquark Fock component  of light scalar mesons are quenched except for \cite{tetra_dyn_proc,tetra_dyn}. All take tetraquark source/sink and omit the disconnected contractions in order to look for states with four valence quarks. The disconnected diagrams are omitted also since they are expensive for numerical evaluation and since they are often noisy.    

\begin{itemize}
\item Prelovsek {\it et al.} performed the $N_f=2$ dynamical and quenched simulation and extracted three lowest energy states in non-exotic $I=0,1/2$  and the exotic $I=2,3/2$ channels using the variational method and a number of $[\bar q\bar q][qq]$ and $(\bar qq)(\bar qq)$ interpolators \cite{tetra_dyn_proc,tetra_dyn}.     
The ground state in all  channels is found to be the scattering state $M_1(0)M_2(0)$ ($M_1M_2=\pi\pi$ or $K\pi$), as demonstrated using the time dependence of the diagonal correlators.  The resulting $Z_i^1(L)$ is also roughly consistent with the expectation for a scattering state $Z_i^1(12)/Z_i^1(16)\simeq (16/12)^{3/2}$  (the ratios for $I=0,1/2$ have sizable errors, which do not allow to make a distinction) \cite{tetra_dyn}.   One of the states in all the channels is close to  $M_1(\tfrac{2\pi}{L})M_2(-\tfrac{2\pi}{L})$ state. 

Additional light states are found in $I=0$ and $I=1/2$ channels, which may be related to observed $\sigma$ and $\kappa$ resonances with strong tetraquark components. The mass dependence of these candidates for $\sigma/\kappa$ on $m_\pi$ are in  qualitative agreement with prediction of unitarized ChPT \cite{pelaez}. A simulation which takes into account also the disconnected diagrams will be  needed  to verify whether the additional states in $I=0,1/2$ channels are not some kind of unknown artifacts related to the omission of disconnected diagrams. 
In the repulsive $I=2,3/2$ channels no light state in addition to the scattering states $M_1(0)M_2(0)$ and $M_1(\tfrac{2\pi}{L})M_2(-\tfrac{2\pi}{L})$ is found, which is consistent with no experimentally observed resonances in these two channels.   
\item Mathur {\it et al.} extract three lowest states in  $I=0$ channel from a single $\pi\pi$ correlator using the sequential Bayes method  \cite{mathur_scalar} and a quenched simulation.   The ground state energy is consistent with $\pi(0)\pi(0)$ and its coupling is consistent with scattering state $Z^1(12)/Z^1(16)\simeq (16/12)^{3/2}$. The energy of the third state is consistent with $\pi(\tfrac{2\pi}{L})\pi(-\tfrac{2\pi}{L})$. They find an additional state in between, which behaves according to a one-particle expectation  $Z^1(12)/Z^1(16)\simeq 1$. This state is a candidate for the observed $\sigma$ resonance with a strong tetraquark component. The presence of an additional state  needs to be confirmed by a simulation that takes into account the disconnected contractions.
\item Suganuma {\it et al.}  extract  the ground state from a single $[\bar q\bar q][qq]$ correlator in $I=0$ channel \cite{suganuma}. They employ the conventional and hybrid boundary conditions which indicate that their ground state is a $\pi\pi$ scattering state.
\item Alford and Jaffe extract  the $I=0,2$ ground state energy $E_1(L)$ from a $\pi\pi$ correlator for a number of lattice sizes $L$ \cite{jaffe_lat}. They argue that $E_1^{I=2}(L)$  is in accordance with a scattering state, while $E_1^{I=0}(L)$ departs from the expected behavior for scattering states and may be an  indication for $\sigma$.   
\end{itemize} 

\section{Hidden charm resonances}
The most prominent tetraquark candidate is the charged $Z^+(4430)$ resonance, discovered by Belle \cite{Zbelle}: it decays to $\pi^+\psi^\prime$, so it must have a minimal quark content $\bar du\bar c c$, but it has not been confirmed by Babar  \cite{Zbabar}. I will also discuss the observed neutral hidden charmonium resonances $X(3872)$, $Y(4260)$ and $Y(4140)$ \cite{swanson,pdg},  which are candidates for tetraquarks or mesonic molecules, although here the charmonium $\bar cc$ Fock component can not be straightforwardly excluded based on the charge alone.

\begin{itemize}
\item Chiu and Hsieh \cite{hsieh} simulated states $\bar c \bar q c q$, $\bar c \bar s c s$, $\bar c \bar c c c$ and  $\bar c \bar q c s$ ($q=u,d$) with $J^{PC}=1^{++}$ and $J^{PC}=1^{--}$. 
They used quenched simulation with overlap valence quarks and omit the 
disconnected diagrams. They extracted only the ground state energy $E_1$ and coupling $Z^1(L)$ from diagonal correlator $C_{ii}(t)$   at two different $L=20,24$. 

The ground $\bar c \bar q c q$ state  with $J^{PC}=1^{++}$ is found at $3890\pm 30$ MeV, which is indeed close to the mass of the $X(3872)$. They find $Z^1(20)/Z^1(24)\simeq 1$, indicating a one-particle (tetraquark/molecular) state. Note however, that the lowest $DD^*$ S-wave scattering state with $E=m_D+m_{D*}\simeq 3879$ MeV  is extremely close and that it should be found in addition to the one-particle state before the indication for the tetraquarks/molecules can be fully trusted.

 The $\bar c \bar s c s$  state  with $J^{PC}=1^{++}$ was found (predicted) at $4100\pm 50$ MeV and $Z^1(20)/Z^1(24)\simeq 1$.  A state with similar properties $Y(4140)$ was indeed later observed by CDF \cite{cdf}. The state is again very close to the scattering threshold $m_{\phi}+m_{J/\psi}\simeq 4117$ MeV, so the scattering state has to be found also in order to trust the existence of tetraquark/molecule.

 The ground $\bar c \bar q c q$ state  with $J^{PC}=1^{--}$ is found at $4238\pm 31$ MeV, which is indeed close to mass of the $Y(4260)$. They find $Z^1(20)/Z^1(24)\simeq 1$, indicating a one-particle (tetraquark/molecular) state.

In all three cases a one-particle nature was deduced from $Z^1(20)/Z^1(24)\simeq 1$. However,  the cautionary remarks concerning $Z(L)$ \cite{alexandrou} have to be kept in mind before concluding that tetraquark/molecule really exist.  
\item The quenched simulation \cite{Z_chinese} was to my knowledge the only one aimed at the very interesting state $Z^+(4430)$. The quantum numbers of these state are not established experimentally, but since it is very close to the $D_1D^*$ threshold, the simulation  \cite{Z_chinese} is carried out in   $J^P=0^-,~1^-,~2^-$ channels. The scattering lengths $a$ are extracted with the help of asymmetric box $L_1\times L_2\times L_3\times T$, which allows for a variety of spatial momenta $k_i=\tfrac{2\pi}{L_i}$. The most reliable results are obtained for 
$J^P=0^-$, where the attractive interaction  between $D_1$  and $D^*$ is found and $a>0$. But the $a$ does not change sign with falling $m_\pi$ and the authors conclude that the attraction is probably to weak to form a loosely bound state.   

\item Liuming Liu  determined a number of S-wave scattering lengths $a$ for scattering between heavy-light, heavy-heavy and light-light mesons using $2+1$ dynamical simulation \cite{liuming}. She studies only channels where no disconnected diagrams are present. 
The scattering lengths are determined from energy shifts $\Delta E=E_1-m_{M_1}-m_{M_2}$ and  the ground state energies $E_1$ are obtained using the   $M_1M_2$ interpolators.  As far as tetraquarks/molecules  are considered, the most interesting result comes from the $D^+\bar D^{0*}$ channel with $I=1$, where $a$ changes sign at $m_\pi\simeq 280$ MeV. This may be an indication for an existence of a loosely bound state at $m_\pi<280$ MeV.   

\item The $2+1$ dynamical simulation of Ehmann and Bali  \cite{ehmann} was actually not aimed at searching for tetraquarks/molecules but to study the mixing between charmonia ($J/\psi$, $\eta_c$, $\chi_c$) and $D\bar D$ ($D$ stands for $D$, $D^*$, $D_1$) states. They compute the a full correlation matrix with $\bar c\Gamma c$  as well as  $(\bar c\Gamma q)(\bar q \Gamma c)$ interpolators in  $J^{PC}=0^{-+},1^{--},2^{++}$ channels, taking into account also all   disconnected contractions. Using the  variational method they determine the spectrum and also the components $\bar c\Gamma c$  and $(\bar c\Gamma q)(\bar q \Gamma c)$ of the  charmonia and  $D\bar D$ physical eigenstates. They do not find any states in addition to the expected charmonia and scattering states and they do not attempt to establish whether their resulting states are one-particle or scattering states. Let me note that a nonzero  coupling $\langle0|(\bar c\Gamma q)(\bar q \Gamma c)|J/\psi\rangle$, for example,   does not mean that $J/\psi$ has a sizable  tetraquark component, since $(\bar c\Gamma q)(\bar q \Gamma c)$ and $\bar c\Gamma c$ Fock components  mix via singly disconnected contractions in \cite{ehmann}.

\end{itemize}

\section{Conclusions }

Proving a sizable tetraquark or molecular Fock component in a hadronic resonance using lattice QCD simulation is not an easy task. A resonance appears as a state in addition to the discrete tower of scattering states. So the extraction of few  states in addition to the ground state is expected to be crucial.  
Given the resulting physical eigenstates, one needs to determine whether a certain state corresponds to a one-particle (tetraquark/molecular) or a two-particle (scattering) state, and  the available methods  to distinguish both  are reviewed.  

There are some indications for an additional state in $I=0,~1/2$ light scalar channels, which might correspond to observed $\sigma$ and $\kappa$ with strong tetraquark components \cite{tetra_dyn_proc,tetra_dyn,mathur_scalar}. The corresponding simulations omitted the disconnected contractions when calculating correlators with tetraquark interpolators  in order to study genuine tetraquark states with four valence quarks. It would be valuable to verify in the future whether the additional states are present also in  a simulation which takes into account the disconnected contractions.

There have been surprisingly few lattice simulation of very interesting exotic $XYZ$ resonances, discovered recently in B-factories. Most of simulations extract only the ground state in a given channel and then try to determine whether it corresponds to a one-particle (tetraquark/molecular) state or to a scattering state. There is some indication that $X(3872)$, $Y(4260)$ and $Y(4140)$ are tetraquark/molecular  states, but future simulation are needed to verify that. \\     
 

\newpage

{\bf Acknowledgments:} I would like to thank my collaborators in the study \cite{tetra_dyn_proc,tetra_dyn} T. Draper, C.B. Lang, M. Limmer, K.-F. Liu, D. Mohler and  N. Mathur  as well as  T.W. Chiu, C. Ehmann, R. Edwards, G. Engel,  C. Gattringer,  J. Juge,  M. Komelj, C. Morningstar, C. Liu, L. Liu, K. Orginos,  J. Pelaez  and S. Sasaki for illuminating discussions.  This work is supported by the Slovenian Research Agency, by the European RTN network FLAVIAnet (contract number MRTN-CT-035482) and  by the Slovenian-Austrian bilateral project (contract number  BI-AT/09-10-012).

\end{document}